\documentclass[12pt]{article}
\usepackage{graphicx}
\begin{document}

\title{Statistical and dynamical aspects in the decay 
of hot neutron-rich nuclei}

\author{
M. Veselsk\'y$^1$, A.L. Keksis$^2$, G.A. Souliotis$^2$, K. Wang$^3$, \\ 
E. Bell$^2$, D.V. Shetty$^2$, M. Jandel$^2$, S.J. Yennello$^2$, Y.G. Ma$^3$ \\
\\
\footnotesize 
$^1$ Institute of Physics, Slovak Academy of Sciences, Bratislava, Slovakia \\
\footnotesize 
$^2$ Cyclotron Institute, Texas A\&M University, College Station, USA \\
\footnotesize 
$^3$ Shanghai Institute of Applied Physics, Shanghai, China
\normalsize
}

\date{}

\maketitle

\begin{abstract}
A signal of isospin-asymmetric phase transition 
in the evolution of the chemical potential was observed 
in hot quasi-projectiles produced in the reactions $^{40,48}$Ca + $^{27}$Al 
confirming an analogous observation 
in the lighter quasi-projectiles observed in the reaction 
$^{28}$Si + $^{112,124}$Sn \cite{SiSnIso}. 
With the increasing mass, the properties of hot quasi-projectiles 
become increasingly influenced by secondary emission. 
Thermodynamical observables exhibit no sensitivity to a different 
number of missing neutrons in the two reactions $^{40,48}$Ca + $^{27}$Al 
and provide a signal of dynamical emission of neutrons, 
which can be related to a very neutron-rich 
low-density region ( neck ) between the projectile and target. 
\end{abstract}

\section*{Introduction}

The isotopic composition of nuclear reaction products provides 
an important information on reaction dynamics and on possible occurrence of 
a phase transition in the isospin asymmetric nuclear matter 
\cite{Serot,KolSanzh}, 
which is supposed to separate into a symmetric dense 
phase and an asymmetric dilute phase. It has been discussed in the literature 
\cite{Spinod} to what extent is such a phase transition 
generated by fluctuations of density or concentration, typically 
suggesting a coupling of both instability modes. 
The N/Z (neutron to proton ratio) 
degree of freedom and its equilibration was studied experimentally 
in detailed measurements of the isotopic distributions of emitted fragments 
\cite{SJY1,Johnston,Ram,RLSiSn,Xu}. Isotopically resolved data 
in the region of Z=2--8 revealed systematic trends, which were, however, 
substantially affected by the decay of the excited primary fragments. 

In order to investigate thermodynamical properties of the 
hot multifragmentation source, it has to be properly characterized 
in terms of the production mechanism and the level of equilibration. 
The fragment data from the reactions 
$^{28}$Si+$^{112,124}$Sn at projectile energy of 30 and 50 MeV/nucleon  
\cite{SiSnIso,SiSnNExch,IsoDist,MuTemp,MVCorrSig} provided full information 
(with the exception of emitted neutrons) on the decay of thermally 
equilibrated hot quasi-projectiles 
with known mass (A=20--30), charge, velocity and excitation energy. 
An excellent description of both the 
dynamical properties of the reconstructed quasi-projectile such as 
velocity, excitation energy and isospin-asymmetry as well as the fragment 
observables 
such as multiplicity, charge and isotope distributions was obtained  
\cite{SiSnNExch} using the model of deep-inelastic transfer (DIT) 
\cite{DITTGSt} for the early stage of collisions, and 
the statistical multifragmentation model (SMM) \cite{SMM} for de-excitation. 
The contribution from non-equilibrium processes 
such as pre-equilibrium emission was shown to be weak \cite{SiSnNExch}. 
The model calculation shown that the average number of neutrons not detected 
in the experiment was between one and two per event. Therefore,    
no significant distortion of the results can be expected. 
The observed trends of thermodynamical observables provided 
several correlated signals of isospin-asymmetric liquid-gas phase 
transition \cite{SiSnIso,MVCorrSig}, in particular a unique signal 
seen in the evolution of the isovector chemical potential \cite{SiSnIso}. 

The present work is an extension of these studies, with the aim to 
investigate the effect of the mass of hot nuclei on observed 
thermodynamical properties and related signals of isospin-asymmetric 
liquid-gas phase transition.

\section*{Experiment}

The experiment was performed at the Cyclotron Institute of 
Texas A\&M University, using 45 MeV/nucleon $^{40,48}$Ca beams delivered by the K500 
superconducting cyclotron impinging on Al (2.2 mg/cm$^2$) target. 
The multidetector array FAUST \cite{FAUST} was employed covering forward 
angles between 
2 and 35 $^{\circ}$, 
where fragments originating from the projectile-like source 
can be expected due to beam velocity. FAUST consists of 68 charged particle 
telescopes arranged in five rings.  Each particle telescope 
consists of a 300 $\mu$m thick silicon detector followed by a 3 cm thick CsI(Tl) 
crystal. The angular ranges covered by the rings were chosen in order to 
distribute the multiplicity of detected particles evenly, thus avoiding cases 
where one telescope is hit simultaneously by multiple particles. 
In the experiment, the mass and atomic number 
of the detected charged particles were identified up to Z=8, for 
the detectors in the forward rings. 

The method of isotope identification 
uses a particle telescope technique in which the isotopes 
are resolved in the two-dimensional $\Delta$E-E spectra.
We used a method \cite{MVCal} which enables to perform the isotope 
identification and energy calibrations simultaneously using a minimization 
procedure. In the experimental spectra, 
the lines for three known isotopes (typically $^{1}$H, $^{4}$He, $^{7}$Be) 
are assigned and energy calibration 
is performed by the minimization procedure where these lines are fitted 
to corresponding calculated energy losses. 
The calibration coefficients are thus obtained as optimum values 
of the minimization parameters. The silicon detectors were calibrated using 
an alpha-source while the empirical formula of Tassan-Got \cite{TGCal} was used 
for the energy calibration of the CsI(Tl) crystals. 

\section*{Analysis}

Using the calibration and identification procedures described above, 
it was possible to identify the charged particles and to determine their 
energies on an event-by-event basis. For the subset of events where all 
detected charged particles were identified, it was possible to 
reconstruct the mass, charge and excitation energy of the composite 
system, in the same way as in the previous work \cite{SiSnNExch}. 
The analysis was performed on the subset of events with total charge larger 
than that of the projectile (Z$\ge$21), thus selecting events where 
the incomplete fusion, 
occurring in mid-central collisions, is the dominant contributing reaction 
mechanism. 

\subsection*{Statistical decay of the hot source with mass 40 -- 50}

One of the goals of the present work was to verify the behavior observed 
in the reaction $^{28}$Si + $^{112,124}$Sn. Of special interest was the 
verification of the signal 
of the liquid-gas phase transition, obtained using isoscaling analysis 
in our previous work \cite{SiSnIso}.  
Isoscaling \cite{TsangIso} is observed when the ratio of isotopic yields 
from two reactions with a different isospin asymmetry  
exhibits an exponential dependence on the fragment isospin asymmetry 

\begin{equation}
     R_{21}(N,Z) = Y_{2}(N,Z)/Y_{1}(N,Z) \simeq C \exp( \beta^{\prime} (N-Z) )
\hbox{ ,}
\label{r21isobt}
\end {equation}

\noindent
where the parameter $\beta^{\prime}$ can be related, in the 
grand-canonical limit, to the isovector 
component of the free nucleon chemical potential, since
$\beta^{\prime}$ = $\Delta (\mu_{n}-\mu_{p})$/2T.

The fragment data obtained in the reactions
$^{28}$Si+$^{112,124}$Sn at projectile energy 30 and 50 MeV/nucleon
\cite{SiSnIso,SiSnNExch,IsoDist,MuTemp} provided full information
(with the exception of emitted neutrons) on the decay of thermally
equilibrated hot quasi-projectiles
with known mass (A=20--30), charge, velocity and excitation energy.
A simulation employing the model of deep-inelastic transfer (DIT)
\cite{DITTGSt} for the early stage of collisions and
the statistical multifragmentation model (SMM) \cite{SMM} for de-excitation 
allowed excellent description of both the
dynamical properties of the reconstructed quasi-projectile such as
velocity, excitation energy and isospin-asymmetry and the fragment observables
such as multiplicity, charge and isotope distributions \cite{SiSnNExch}. 
Thermodynamical properties 
of the quasi-projectiles undergoing statistical
multifragmentation \cite{SiSnIso,SiSnNExch,IsoDist,MuTemp} were investigated 
in order to reveal possible signals of the phase transition.
The difference of isovector chemical potentials in the two reactions
$\Delta(\mu_n-\mu_p)$ was estimated using isoscaling analysis.  
A turning-point in the trend of the observable $\beta^{\prime} T$ 
was observed at excitation energy of about 4 MeV/nucleon. Such a behavior 
can be understood as a signal of the onset
of chemical separation into the dense isospin symmetric and 
the dilute isospin asymmetric phases which reverts the decreasing trend  
of the free nucleon
chemical potential consistent with expansion of the homogeneous system. 
Above the turning-point, the temperature determined using the fragment 
yield ratios agreed well with the kinematic temperature of protons, 
which can be identified as the remnants of the dilute phase (a nucleon gas). 
Such agreement between the two methods of thermometry demonstrates 
that quantitatively, the grand-canonical 
approach describes the properties of 
hot nuclei undergoing multifragmentation very well. 
Besides isoscaling, another grand-canonical scaling  
was observed since the isobaric yield ratio Y($^{3}$H)/Y($^{3}$He) 
exhibited an exponential dependence on the quasiprojectile N/Z ratio. 
The experimental data from reactions $^{28}$Si+$^{112,124}$Sn 
at projectile energy 30 and 50 MeV/nucleon are thus well understood 
in terms of a reaction mechanism, where the projectile and target 
nuclei form a di-nuclear configuration, exchange a considerable amount 
of nucleons, and after the re-separation reach thermal equilibrium 
at different temperatures. The highly excited quasi-projectile 
continues to expand up to the spinodal contour. 

\begin{figure}[h]
\centering
\vspace{5mm}
\includegraphics[width=11.0cm,height=9.cm]{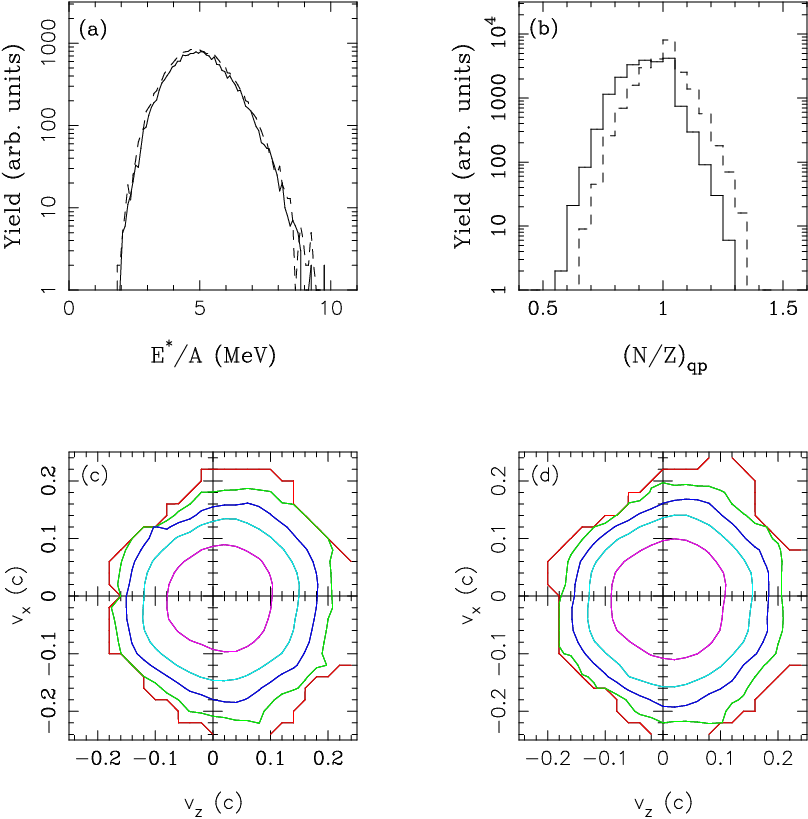}
\caption{\footnotesize  
(a) Excitation energy distributions of the reconstructed quasi-projectiles 
in the reactions $^{40,48}$Ca + $^{27}$Al at 45 MeV/nucleon 
(solid line for $^{40}$Ca and dashed line for $^{48}$Ca),  
(b) N/Z-distributions of the reconstructed quasi-projectiles 
(solid line for $^{40}$Ca and dashed line for $^{48}$Ca),  
(c) Velocity-plot of the light charged particles in the 
reaction $^{40}$Ca + $^{27}$Al at 45 MeV/nucleon, 
(d) Velocity-plot of the light charged particles in the 
reaction $^{48}$Ca + $^{27}$Al at 45 MeV/nucleon. 
}
\label{qpchar}
\end{figure}

\begin{figure}[h]
\centering
\vspace{5mm}
\includegraphics[width=11.0cm,height=9.cm]{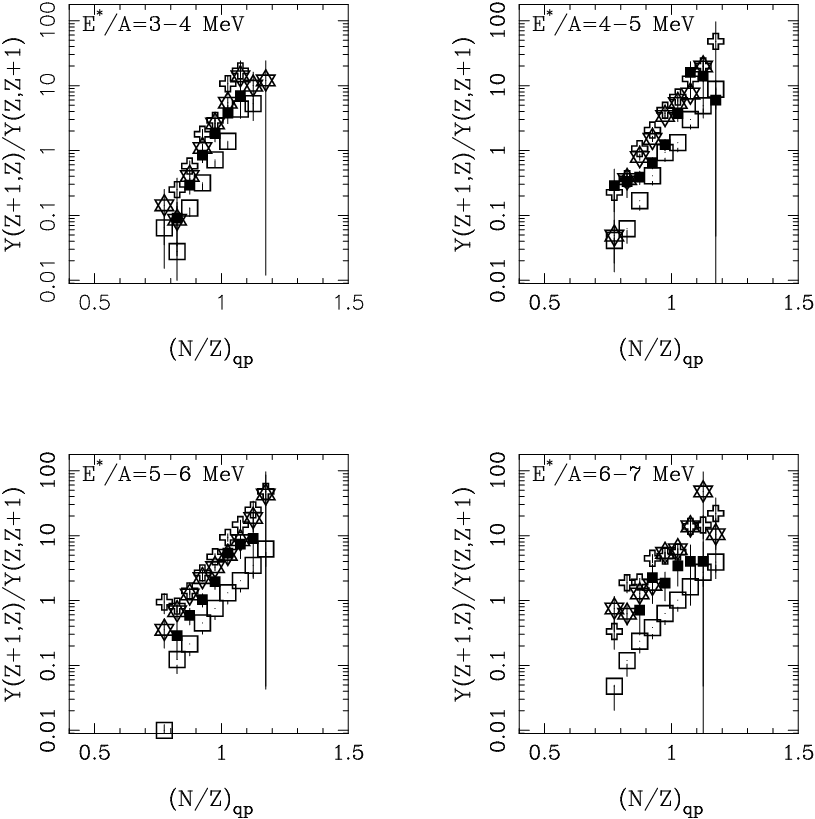}
\caption{\footnotesize  
Yield ratios of mirror nuclei ( $^{3}$H/$^{3}$He - open squares, 
$^{7}$Li/$^{7}$Be - thick crosses, $^{11}$B/$^{11}$C - stars, 
$^{15}$N/$^{15}$O - solid squares )  
observed in the reaction $^{48}$Ca + $^{27}$Al at 45 MeV/nucleon 
plotted for four excitation energy bins as a function of N/Z 
of the reconstructed projectile-like nuclei. 
}
\label{isbscal}
\end{figure}

As a first step of the analysis, the characteristics of the projectile-like 
nuclei, observed at forward angles, were reconstructed on 
an event-by-event basis. 
The results for the fully resolved quasiprojectiles with Z$\ge$21, which 
are expected to originate dominantly from incomplete fusion reactions 
at mid-central impact parameters, are shown in Fig. \ref{qpchar}. 
The excitation energy distributions of the reconstructed quasi-projectiles 
in the reactions $^{40,48}$Ca + $^{27}$Al at 45 MeV/nucleon 
(see Fig. \ref{qpchar}a) are practically identical, what is somewhat 
surprising when taking into account that the number of missing 
(undetected) neutrons may differ considerably. A possible uncertainty 
in the evolution of neutrons can be documented 
also by the observed N/Z-distributions (see Fig. \ref{qpchar}b) which reflect 
initial N/Z-difference of the two projectile nuclei (amounting to 0.48) 
only partially, since the 
mean values (centroids) in the reactions $^{40,48}$Ca + $^{27}$Al 
are N/Z=0.93 and 1.01, respectively. The velocity-plots of the light charged 
particles in the quasi-projectile frame in these 
two reactions are shown in Fig. \ref{qpchar} (for $^{40}$Ca 
on panel \ref{qpchar}c and for $^{48}$Ca on panel \ref{qpchar}d).  
Practically isotropic emission is observed which again implies 
the statistical multifragmentation of the 
hot projectile-like nucleus. A slight suppression at the backward 
hemisphere is caused by the combined effect of limited angular 
coverage and energy thresholds of the experimental device.    
Using the observed characteristics of the quasi-projectile, 
the centrality of the observed data was estimated using the simulation, 
successful in describing properties of the hot multifragmenting 
sources formed in violent collisions in many reactions in the Fermi 
energy domain \cite{MVProd,MVSnAl}. Comparison of the 
properties of the hot quasiprojectile source leads to conclusion 
that incomplete fusion collisions at mid-central 
impact parameters contribute dominantly, mainly due to narrow 
angular acceptance of the experimental setup around the beam direction 
for the reconstructed quasi-projectiles with Z$\ge$21. Similar 
selectivity was experimentally observed in heavy residue data 
in the reaction $^{124}$Sn+$^{27}$Al \cite{MVSnAl}.

\begin{figure}[h]
\centering
\vspace{5mm}
\includegraphics[width=13.0cm,height=8.cm]{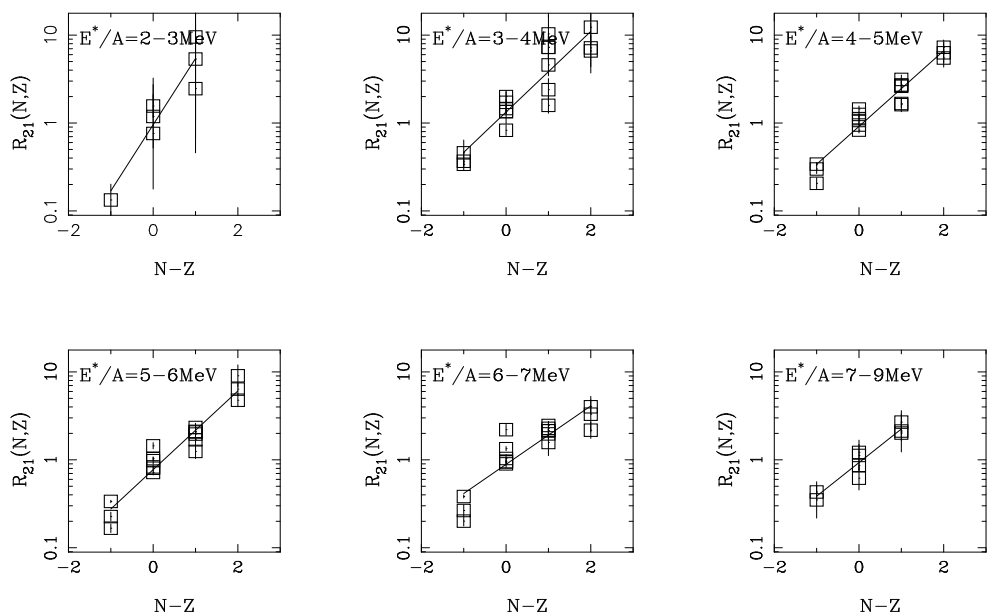}
\caption{\footnotesize  
Isoscaling plots, constructed for the pair of reactions 
$^{40,48}$Ca + $^{27}$Al at 45 MeV/nucleon using the identified 
fragments with Z=1-8 forming the fully resolved 
quasiprojectiles with Z$\ge$21 in six excitation energy bins. 
}
\label{fisosc}
\end{figure}

Statistical emission in the projectile-like frame is indicated 
also in Fig. \ref{isbscal}. In a similar way as in the reactions 
$^{28}$Si+$^{112,124}$Sn, where the isobaric ratio $^{3}$H/$^{3}$He exhibited 
an exponential (grand-canonical) scaling with the quasiprojectile N/Z, 
several yield ratios of mirror nuclei measured in the reaction 
$^{48}$Ca + $^{27}$Al at 45 MeV/nucleon are plotted as functions of the N/Z 
of the reconstructed projectile-like nuclei. Again the exponential 
scaling is observed, with identical slopes for all ratios within each 
of the excitation energy bins. The slope decreases with increasing 
excitation energy in an analogous way as in the reactions 
$^{28}$Si+$^{112,124}$Sn \cite{MuTemp}.  
The observed grand-canonical scaling provides an evidence 
that statistical multifragmentation 
is a dominant mode of de-excitation also in the present case.   
It is also remarkable to note that essentially no effect of missing 
neutrons on the observed slope can be seen in this neutron-rich 
system with a significant number of missing neutrons. 
  
\begin{figure}[h]
\centering
\vspace{5mm}
\includegraphics[width=9.0cm,height=7.5cm]{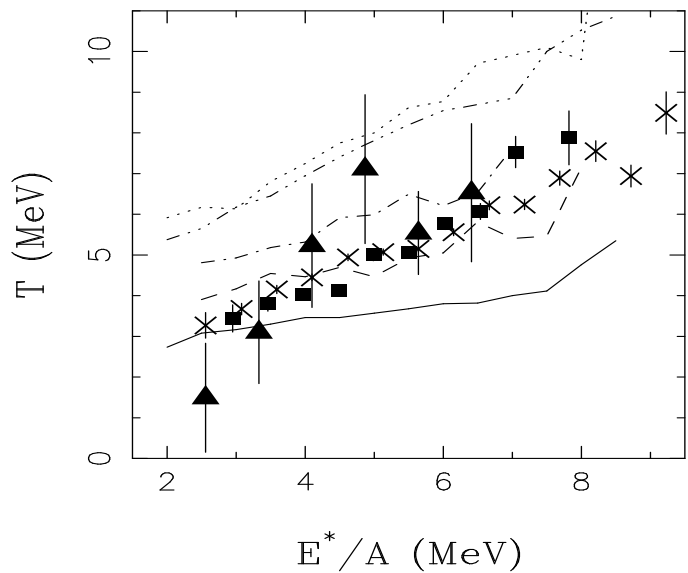}
\caption{\footnotesize  
Caloric curve obtained 
using the double isotope ratio thermometer d,t/$^{3,4}$He 
(squares), compared to the thermometer d,t/$^{6,7}$Li (triangles), 
and to the kinematic temperatures of protons, deuterons, tritons, 
$^{3}$He and alpha particles (solid, dashed, single dot-dashed, dotted 
and multiple dot-dashed lines, respectively). 
Combined fully resolved data with Z$\ge$21 
from two reactions $^{40,48}$Ca + $^{27}$Al at 45 MeV/nucleon was used.  
Thin crosses show caloric curve obtained using the thermometer d,t/$^{3,4}$He 
in the reactions $^{28}$Si+$^{112,124}$Sn \cite{MuTemp}.
}
\label{ftempdt34he}
\end{figure}

As in the previous work \cite{SiSnIso}, the isoscaling analysis was performed 
also for the reactions  $^{40,48}$Ca + $^{27}$Al at 45 MeV/nucleon. 
Figure \ref{fisosc} shows isoscaling plots, plotted as a function 
of the projection of isospin, constructed using the identified 
fragments for the fully resolved quasiprojectiles with  Z$\ge$21 
in six excitation energy bins. The observed isoscaling behavior is 
quite regular and the value of the isoscaling parameter (slope) decreases 
with excitation energy, in agreement with the previous experiment 
\cite{SiSnIso} and other experiments reported in the literature.

\begin{figure}[h]
\centering
\vspace{5mm}
\includegraphics[width=13.0cm,height=11.cm]{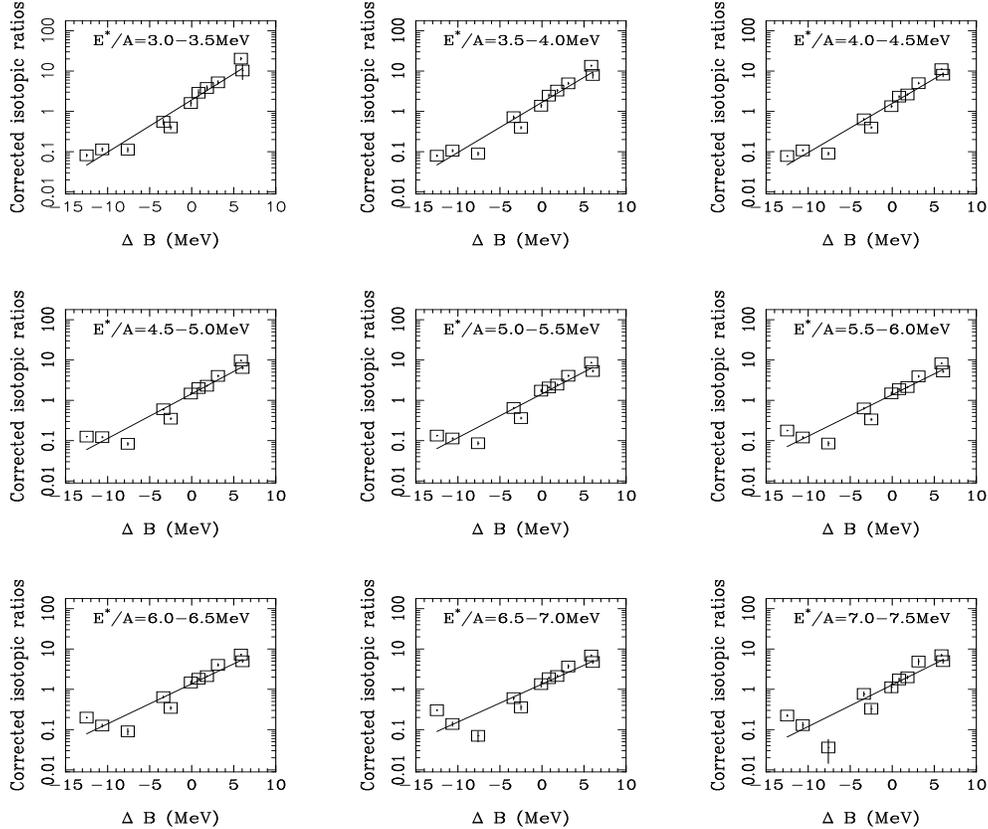}
\caption{\footnotesize  
Isotopic yield 
ratios, corrected for the mass and for the ground state spin, 
plotted as a function 
of difference of binding energies for nine excitation energy bins. 
}
\label{fyrtemp1a}
\end{figure}

In order to estimate the evolution of the chemical potential, 
as in the previous work \cite{SiSnIso}, one needs to estimate 
the system temperature. Figure \ref{ftempdt34he} shows the  
caloric curve obtained for the selected fully resolved data with  Z$\ge$21 
using the double isotope ratio thermometer d,t/$^{3,4}$He (squares). 
Furthermore, Figure \ref{ftempdt34he} shows the comparison 
of the double isotope ratio thermometer d,t/$^{3,4}$He to 
another thermometer d,t/$^{6,7}$Li (triangles) and to 
the kinematic (slope) temperatures of protons, deuterons, tritons, 
$^{3}$He and alpha particles. One can see that, unlike the previous 
work \cite{MVCorrSig}, there is no clear correspondence of any slope 
temperature to the double isotope ratio thermometers.  
Thus one needs to establish to what extent the double isotope ratio 
temperature of the two thermometers represents the multifragmentation 
temperature. A good judgment can be obtained when comparing 
the d,t/$^{3,4}$He temperature observed in the present work  
to the earlier results for the reactions $^{28}$Si+$^{112,124}$Sn 
\cite{MuTemp,MVCorrSig}. Such a comparison shows that the obtained 
caloric curves in both reactions are consistent and thus possibly 
represent general properties of the multifragmenting system. On the other 
hand, a similar comparison for the proton kinetic temperature 
shows that in the heavier system it dropped considerably, 
what can be explained by the onset of intense secondary 
emission of protons in the heavier system. Thus the mass range 
observed in the present work can represent an upper limit where 
the effect of secondary emission can be disentangled.  

Moreover, since the double isotope ratio thermometers d,t/$^{3,4}$He 
and d,t/$^{6,7}$Li are just two of many possible thermometers, one can 
define a global temperature as an average value over the larger set 
of double isotope ratio thermometers. 
Such global temperature can be obtained using a graphical method 
developed in our previous works \cite{DANF2001,Shetty}. 
The isotopic yield 
ratios, corrected for mass and ground state spin, can be plotted as a function 
of the difference of binding energies. The resulting plots for the 
present data are shown in Figure \ref{fyrtemp1a} for 
nine excitation energy bins. 

\begin{figure}[h]
\centering
\vspace{5mm}
\includegraphics[width=13.5cm,height=6.5cm]{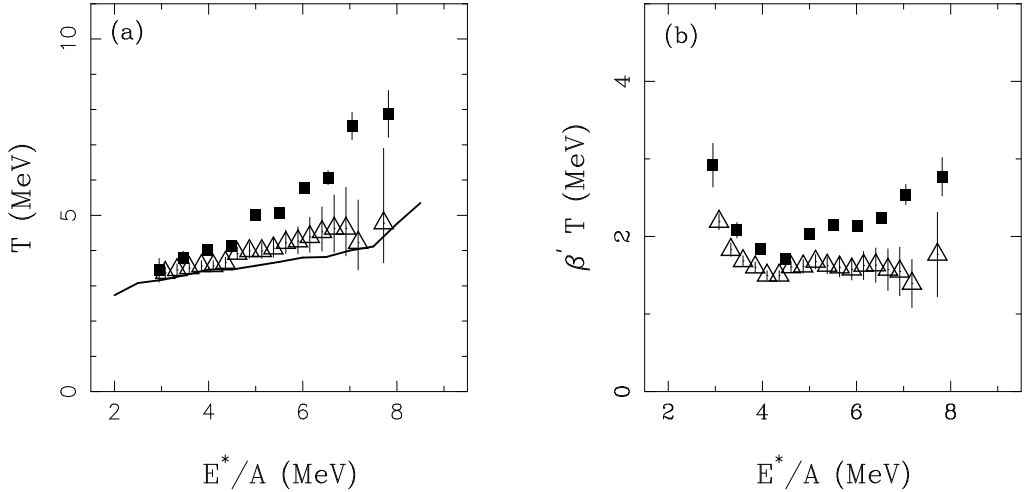}
\caption{\footnotesize  
(a) Caloric curve obtained using global temperature 
from the slopes of $\Delta B$ dependences 
in Fig. \ref{fyrtemp1a} ( triangles ) along with the caloric curves 
obtained using the double isotope ratio thermometer 
d,t/$^{3,4}$He ( squares ) and proton slope temperatures 
( line ). 
(b) Resulting values of $\beta\prime T$ obtained using 
the global ( triangles ) and d,t/$^{3,4}$He 
( squares ) temperatures. 
}
\label{fyrtemp}
\end{figure}

The observed exponential (and thus again grand-canonical) scaling 
is reasonably good and one can assume that the 
fitted slope provides an global double isotope ratio temperature. 
Figure \ref{fyrtemp}a shows the resulting caloric curve, 
along with the results of the double isotope ratio thermometer 
d,t/$^{3,4}$He and the kinematic temperatures. It is remarkable to note, that 
there is a good correspondence between the global isotope ratio 
temperature and proton kinematic temperature. Both curves are relatively 
flat above 4 AMeV, which can be possibly interpreted as a long plateau. 
However, comparison with the lighter hot system in ref. \cite{SiSnIso} shows  
that the value of the plateau temperature is much lower than it can expected 
in multifragmentation. Especially in the case of thermometers using 
proton multiplicity and poton kinamatic temperature 
such low temperature with flat behavior can be caused by increased 
influence of the secondary de-excitation via nucleon (proton) emission, 
Such mode is not dominant in the lighter system,in ref. \cite{SiSnIso} 
where the Fermi decay, analogous to multifragmentation, 
dominates. Thus the proton kinematic temperature for the lighter system 
represents the earlier de-excitation stage, as is documented by higher 
values of the temperature \cite{MVCorrSig} than in the present case.  

The correspondence of the global isotope ratio
temperature and the proton kinematic temperature can thus be explained 
by secondary emission. The scaling behavior in Fig. \ref{fyrtemp1a} 
is determined mainly by isotopic ratios of intermediate mass fragments up to 
oxygen, where the nucleon (proton) emission is an increasingly 
dominant mode of secondary de-excitation. Thus, it is not  
surprising that identical temperatures are extracted from both the 
Maxwellian spectra and the fragment yield ratios.  
Unlike proton emission, the surface emission 
of the isotopes used in the d,t/$^{3,4}$He thermometer is less probable.  
Therefore, their properties (as seen in their kinematic temperatures) 
reflect multifragmentation (volume emission) more closely. 
This is supported by the fact that the the isotopic ratio $^{3}$He/$^{4}$He 
(first point from the left in each panel of Fig. \ref{fyrtemp1a}) 
does not follow the systematics. 
The other isotopic ratio falling out of the systematics 
appears to be $^{15}$O/$^{16}$O ( third from the left ). 
Its deviation from the systematics becomes larger with increasing 
excitation energy, and can be caused by an 
increasing influence of low lying excited states.  
A spin degeneracy factor of these excited states can be larger than the one 
of the ground state which was assumed in the correction.    
One thus ends up with the two distinct groups of thermometers. In the first 
group are the average isotope ratio
temperature and proton kinematic temperature ( and also
the double isotope ratio thermometers including protons ), 
representing the stage of secondary emission. The second group, 
the double isotope ratio thermometers d,t/$^{3,4}$He
and d,t/$^{6,7}$Li, appear to represent 
an earlier stage of de-excitation, i.e., multifragmentation of hot nuclei. 
The kinematic temperatures of $d$,$t$,$^{3,4}$He also appear to be consistent 
with the earlier stage, as was the case in the previous work \cite{MVCorrSig}.  
However, especially for deuterons and tritons one can not exclude admixture 
from the later stage, either by emission or by coalescence. 

In macrocanonical limit, a product of the temperature $T$ and 
isoscaling parameter $\beta'$ (from Fig. \ref{fisosc}), 
corresponds to a value of the isovector part of chemical potential. 
The resulting values of $\beta'T$ are shown in Figure \ref{fyrtemp}b 
for the d,t/$^{3,4}$He thermometer (squares) and the average isotope ratio 
($\Delta B$) temperature (from Fig. \ref{fyrtemp1a}, triangles). 
In the chemical potential corresponding to the d,t/$^{3,4}$He thermometer, 
a reversion of the trend is observed at 4 MeV/nucleon, analogous to the results for 
lighter system \cite{SiSnIso}. Such behavior of chemical potential was also 
confirmed by lattice-gas calculations \cite{YGMLG}.  
The estimate of the isovector chemical potential, obtained using the 
thermometer representing the secondary emission, 
still leads to approximately constant behavior above 4 MeV/nucleon. 
It can be concluded that the increase of the isovector chemical potential, 
as documented for the d,t/$^{3,4}$He thermometer, 
is related to de-excitation of the hot multifragmenting source and 
signals increasing isospin asymmetry of the gas phase in the 
isospin-asymmetric liquid-gas phase transition \cite{SiSnIso}. 
Furthermore, one needs to take into account that secondary emission influences 
also the value of the isoscaling parameter. The work in Ref. \cite{IsoEvap} 
showed that such influence is not significant for the lighter system 
( A$\simeq$25 ), 
but for the systems with A$\simeq$50 ( as in the present case ) 
secondary emission and wide initial isotopic distributions from the dynamical 
stage 
of collision result in higher values of isoscaling parameters, and such trend 
weakens with increasing excitation energy. Thus the actual dependence 
can have an even deeper minimum since the effect of secondary emission 
tends to flatten it. 

In general, one can conclude that the trends observed in the lighter 
quasi-projectiles in reactions  $^{28}$Si+$^{112,124}$Sn 
\cite{SiSnIso,MuTemp,MVCorrSig} 
were confirmed in reactions $^{40,48}$Ca + $^{27}$Al at 45 MeV/nucleon 
for the projectile-like nuclei with the mass about twice that of the former. 
Moreover, it has been shown that the properties of the hot systems, 
as reflected by the 
fragment observables, become still more distorted 
by secondary emission with the increasing mass. 

\subsection*{Missing neutrons -- statistical or dynamical emission ?}

Specifically for the neutron-rich projectile $^{48}$Ca, the effect 
of missing (undetected) neutrons must be considered. In the earlier works, 
it was established using successful simulations for the reactions 
$^{28}$Si+$^{112,124}$Sn that the losses due to neutron emission 
represent a relatively small part of the system \cite{SiSnNExch}. The isospin 
equilibration was not complete and thus the neutron-rich target did 
not result in a comparably neutron-rich quasi-projectile. 
The effect of missing neutrons was examined using isoscaling plots 
expressed for sets of identical quasi-projectiles 
from two reactions \cite{SiSnIso}. 
No dependence was observed which could be attributed to the influence 
of missing neutrons. It was also observed that the shape of the 
caloric curve did not depend on the N/Z of the quasi-projectile. 

\begin{figure}[h]
\centering
\vspace{5mm}
\includegraphics[width=9.0cm,height=9.cm]{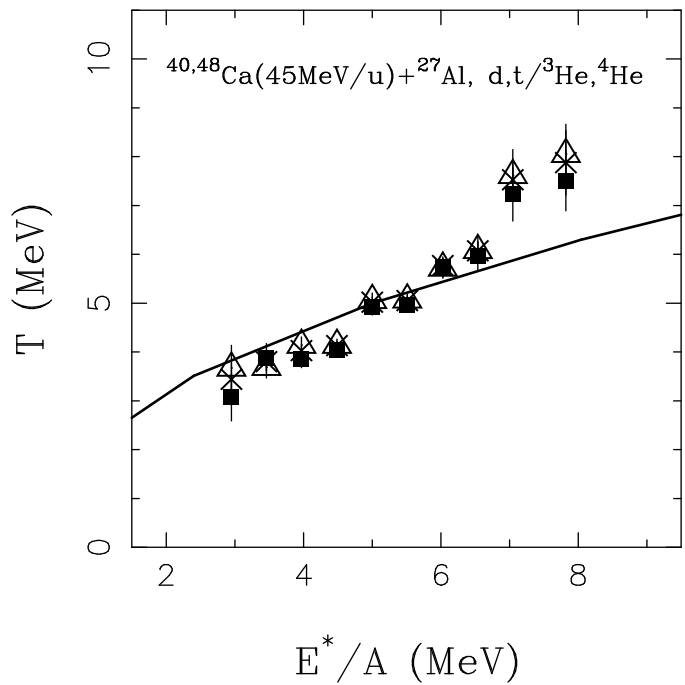}
\caption{\footnotesize  
Caloric curves obtained 
using the double isotope ratio thermometer 
d,t/$^{3,4}$He ( $^{40}$Ca+$^{27}$Al - squares, 
$^{48}$Ca+$^{27}$Al - triangles, combined data - crosses, line - 
theoretical dependence from ref. \cite{PhTrns}  ). }
\label{fcal12}
\end{figure}

In the case of the reactions $^{40,48}$Ca+$^{27}$Al, even when assuming 
an incomplete fusion scenario, one has to expect that there will be 
a considerable difference in neutron excess of the hot systems 
in the two reactions ( seven neutrons according to simulation ). 
In this context, it is remarkable that
the caloric curves, expressed as a function of excitation 
energy obtained by charged particle calorimetry, 
are almost identical in both reactions, despite the expected significant 
difference of total excitation energy after including the missing neutrons.  
Figure \ref{fcal12} shows the caloric curves 
obtained using the double isotope ratio thermometer 
d,t/$^{3,4}$He for the reactions $^{40,48}$Ca+$^{27}$Al, 
along with the theoretical line from ref. \cite{PhTrns}. 
There is a little difference observed between the two reactions.  
Especially, there is no 
shift which could be attributed to the difference of true excitation energies, 
which could exceed 1 MeV/nucleon, due to a different number of emitted neutrons 
in the simulation ( up to seven ). 
The experimental caloric curves agree well with 
the theoretical dependence \cite{PhTrns}, representing temperatures 
where the isolated system with given excitation energy enters the spinodal 
contour, thus confirming the conclusion that, 
as in the reactions $^{28}$Si+$^{112,124}$Sn, 
the d,t/$^{3,4}$He thermometer 
represents multifragmentation of the hot equilibrated source. 
Complementary to the caloric curves, the isoscaling behavior 
for selected N/Z-bins was also verified.  
The observed slopes, representing difference of chemical potentials, 
were consistent with zero, within statistical errors. 
Thermodynamical observables ( see Figs. \ref{qpchar}a, \ref{isbscal} 
and \ref{fcal12} ) thus, remarkably, do not exhibit sensitivity 
to the number of missing ( undetected ) neutrons in two reactions. 

As an explanation one has to assume that the apparent excitation energy, 
reconstructed using observed charged particles, is close to the 
true excitation energy of the hot equilibrated source, and the excess 
neutrons in the neutron-rich case are not part of the equilibrated   
source, due to dynamical emission taking place prior 
to equilibration. 
Such explanation is consistent with almost identical experimental 
excitation energy distributions shown in Fig. \ref{qpchar}a 
for two reactions $^{40,48}$Ca+$^{27}$Al 
and is also supported by the lack of the effect of missing neutrons on 
the scaling shown in Fig. \ref{isbscal} in the neutron-rich 
reaction $^{48}$Ca+$^{27}$Al. 
Dynamical emission of neutron-rich charged particles at mid-velocity 
was reported in recent years \cite{Neck}, and interpreted as 
caused by formation of a neutron-rich neck. 
Such scenario can provide an explanation also in the present case, 
with the notable difference that the neck structure may be formed 
exclusively by neutrons. In the present case the low-density
region between the projectile and target can be expected to be very
neutron-rich, since the nuclear equation of state for sub-saturation
densities provides stable solutions only outside of the spinodal
region and thus at very asymmetric N/Z ratios.
A behavior consistent with the present case was reported 
recently \cite{ColdFrg} in incomplete fusion reactions, 
where an isospin-dependent component of excitation energy  
was necessary to explain the production of neutron-rich fragmentation products.  
Such behavior is consistent with a rupture of the neck structure formed 
by the neutrons 
in the region between the hot and cold pre-fragment during the dynamical stage.
Unlike the reported cases, where signatures of neck formation 
were reported in symmetric damped collisions of massive nuclei \cite{Neck}, 
in the present case the alternative 
explanation due to Coulomb force of the massive external charge \cite{CoulDec}
can be excluded. The remaining 
charge, especially in the incomplete fusion reactions, is rather 
small, and the observed trends can be attributed uniquely to the 
effect of the nuclear mean field. 

\section*{Summary and conclusions}

In summary, the signal of isospin-asymmetric phase transition 
in the evolution of chemical potential was observed 
in the multifragmentation of hot quasi-projectiles in the 
reactions $^{40,48}$Ca+$^{27}$Al confirming 
an analogous observation in lighter quasi-projectiles in the reaction 
$^{28}$Si+$^{112,124}$Sn \cite{SiSnIso}. 
However, with an increasing mass the properties of the hot systems
become increasingly influenced by secondary emission. 
Thermodynamical observables exhibit no sensitivity 
to a different amount of missing neutrons due to variation of neutron excess 
of the hot system in the two reactions $^{40,48}$Ca+$^{27}$Al, 
thus providing a signal of dynamical emission of neutrons, 
which can be related to a formation of a neutron-rich 
low-density region (neck) between the projectile and target. 

The authors wish to thank the staff of the Texas A\&M Cyclotron 
facility for the excellent beam quality. 
This work was supported by the US Department of Energy under contract 
DE-FG03-93ER40773, 
by the Welch Foundation under contract A-1266, by 
the Slovak Scientific Grant Agency under contracts VEGA-2/5098/25 and 
VEGA-2/0073/08, by the Slovak Research and Development
Agency under Contract No. SK-CN-00706 and through the 
Agreement of Scientific Cooperation between China and Slovakia by 
the Ministry of Sciences and Technology and the Major State Basic Research
Development Program (973 program) under contract No. 2007CB815004.

\end{document}